\def\ref{\par\noindent\hang}

\def\spose#1{\hbox to 0pt{#1\hss}}
\def\approxlt{\mathrel{\spose{\lower 3pt\hbox{$\sim$}}
	\raise 2.0pt\hbox{$$<$$}}}
\def\approxgt{\mathrel{\spose{\lower 3pt\hbox{$\sim$}}
	\raise 2.0pt\hbox{$>$}}}

\def\multleft#1{\hbox to size{\vbox {\halign {\lft{##}\cr #1}}\hfill}\par}
\def\multright#1{\hbox to size{\vbox {\halign {\rt{##}\cr #1}}\hfill}\par}

\def\$<${\thinspace}

\def\boxit#1{\vbox{\hrule\hbox{\vrule\kern3pt\vbox{\kern3pt
          #1 \kern3pt}\kern3pt\vrule}\hrule}}

\def\arcm{{\arcm\thinspace arcmin}}

\documentstyle[psfig]{mn}
\begin{document}
\hsize=6truein

\title[]{Coulomb interactions in the intracluster medium}

\author[]
{\parbox[]{6.in} {S. Ettori and A.C. Fabian\\
\footnotesize
Institute of Astronomy, Madingley Road, Cambridge CB3 0HA \\
}}                                            
\date{accepted, October 1997}
\maketitle

\begin{abstract}
In this paper we discuss the effect of Coulomb collisions on the
temperature profiles of the intracluster medium in clusters of galaxies,
motivated by recent reports of negative temperature gradients in some
clusters by Markevitch et al. 
The timescale for electrons and protons to reach
temperature equilibrium can exceed a few billion years beyond radii of a
Mpc, if the intracluster gas is assumed to be at the usual cluster virial
temperature. If a cluster merger has occurred within that time causing the
protons, but not the electrons, to be rapidly heated then a small negative
temperature gradient can result. 
This gradient is larger in clusters with high temperatures and steep
density profiles.

Applying these considerations to the cluster of galaxies A2163, we
conclude that, more plausibly, the observed gradient is due to a lack of
hydrostatic equilibrium following a merger.

\end{abstract}


\section{INTRODUCTION} 

Recent results on steep temperature profiles in some clusters of
galaxies, as measured by {\it ASCA} X-ray spectra (Markevitch 1996 and
Markevitch et al. 1996), have raised questions on the physical condition
of the intracluster medium (ICM). At a radius of a few 100~kpc, the ICM
has a characteristic density $n_{\rm gas} \sim 10^{-3}$ cm$^{-3}$,
temperature $T_{\rm gas}\sim10^8$~K and a heavy element abundance of
about 30 per cent of the solar value. The density drops at larger radii
$r$ approximately as $r^2$.

Assuming a polytropic distribution for the gas, the temperature scales
with density as
\begin{equation}
\frac{T_{\rm gas}}{T_0}=\left( \frac{n_{\rm gas}}{n_0} \right)^{\gamma -1},
\end{equation}
where the polytropic index $\gamma$ ranges between 1 and 5/3, the limits
corresponding to the gas being isothermal and adiabatic, respectively.
Markevitch (1996) found that $\gamma \approx 1.9$ and 1.7 for the
clusters A2163 and A665, respectively ($>1.7$ and 1.3 and the 90 per cent
confidence level). When $\gamma> 5/3$ the gas is convectively unstable,
which tends to produce turbulent motions in the ICM, resulting in a
mixed, homogeneous gas after several sound crossing times. The detection
of a dramatic drop in the temperature profile may be an indication that
those clusters are observed in a very unusual, brief stage in their
existence, perhaps having experienced a major merger within the previous
few billion years.

An alternative possibility to explain the steep temperature decline ,
suggested by Markevitch (1996 and Markevitch et al 1996), is that the
electron temperature, $T_{\rm e}$, which is the quantity measured by X-ray
observations, is not representative of the mean gas temperature, $T_{\rm
gas}$. This can occur, for example if a shock has heated the ICM, with
the protons receiving most of the energy so that the proton temperature
$T_{\rm p}$ greatly exceeds the electron temperature $T_{\rm e}$. Coulomb
scattering then equilibrates the temperatures at a rate (Spitzer 1962):
\begin{equation}
\frac{dT_{\rm e}}{dt}=\frac{T_{\rm p} - T_{\rm e}}{t_{\rm eq}}.
\end{equation}
In this equation, $t_{\rm eq}$, the equipartition time via Coulomb
scattering, is given by
\begin{equation}
t_{\rm eq} = \sqrt{\frac{\pi}{2}} \frac{m_{\rm p}}{m_{\rm e}} 
\frac{\theta_{\rm e}^{3/2}} {n_{\rm p} c \sigma_{\rm T} \ln \! \Lambda},
\end{equation}
where $\theta_{\rm e} = k_{\rm B}T_{\rm e}/(m_{\rm e} c^2)$, 
$n_{\rm p}$ is the proton density,
$\sigma_{\rm T}$ is the Thomson scattering cross section and $\Lambda$ is
the ratio of largest to smallest impact parameters for the collisions
($\ln \! \Lambda \sim 37.8$, Sarazin 1988; note that a similar term in
the corresponding dimensionless proton temperature $\theta_{\rm p}$ is
negligible for the conditions of the ICM). The rate is proportional to
the gas density and so can be low, and the equilibration time long, at
the outer parts of a cluster where the density is least.

Here we examine the equilibration timescale for a typical ICM under the
conservation of  total kinetic energy. 
We find (Section~2) that an implausibly short timescale is required for a
large temperature gradient to be observed. Both the X-ray
emission and the equilibration of the ICM depend upon Coulomb scattering, so
when the rate of emission is high the rate of equilibration is also high.
In Section~3, we apply these considerations to the case of A2163 and
discuss our conclusions in Section~4.

\section{Electron--proton Coulomb interactions}

\begin{figure}
\psfig{figure=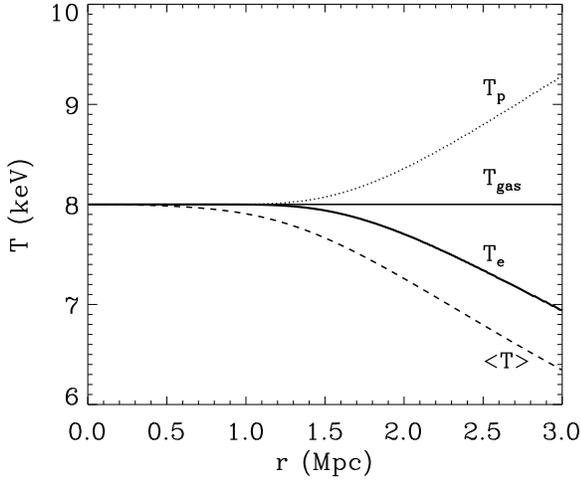,width=.5\textwidth,angle=0}
\caption{ This plot shows 
the temperature profiles as a function of radius $r$ at the time
$t_{\rm m}$ of 3 Gyrs. The input parameters are: $n_0=2.5 \times 10^{-3}
{\rm cm}^{-3}, \beta = 2/3 \ {\rm and} \ r_{\rm c}=0.3$. The thick solid
line indicate the effect of the e--p Coulomb collisions on the electron
temperature
when the energy density $U$ is conserved and assuming that $T_{\rm
gas}(r)$ is isothermal.
The radial decrease of $T_{\rm e}$ with respect to
the central value is about 0.02, 4 and 14 per cent at 1, 2 and 3
Mpc, respectively. 
The dotted line describes the $T_{\rm p}$ profile and the dashed line the
projected temperature $<T>$. 
The latter profile diminishes with respect to the central value by 1, 9
and 21 per cent, at 1, 2 and 3 Mpc, respectively.}
\end{figure}

\begin{figure}
\psfig{figure=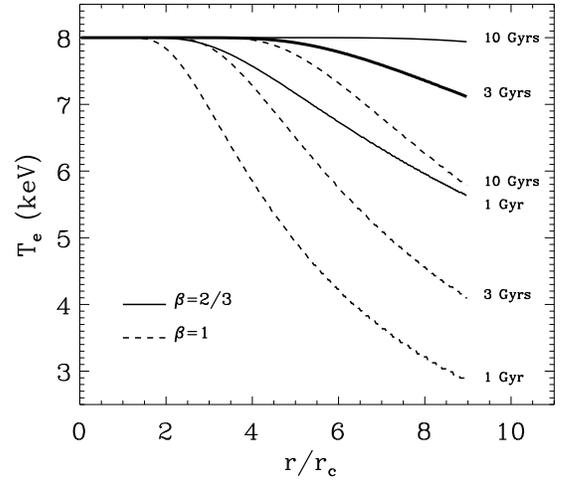,width=0.5\textwidth,angle=0}
\caption{The behavior of the electron temperature as function of
$r/r_{\rm c}$, when $\beta = 2/3$ (solid line) and $\beta = 1$ (dashed
line). The different profile were calculated with different $t_{\rm m}$
(upwards: 1, 3, 10 Gyrs), assuming constant $n_0$ and fixing 
$T_{\rm e}(r=0)$
at 8 keV. The thickest solid line corresponds to $\beta = 2/3$ and
$t_{\rm m} =3$ Gyrs.
} \end{figure}

\begin{figure}
\psfig{figure=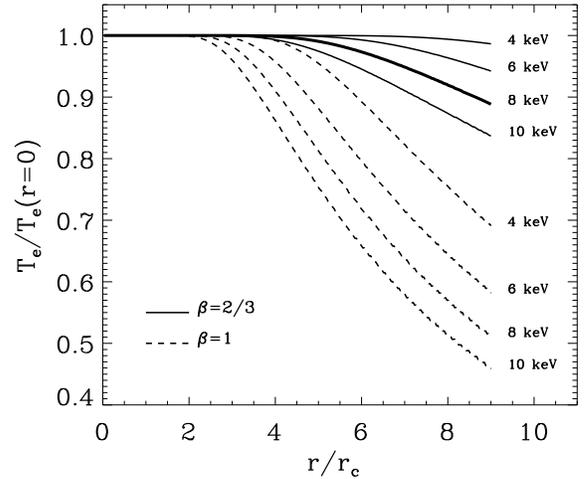,width=0.5\textwidth,angle=0}
\caption{The dependence of $T_{\rm e}$, normalized to the
central value $T_{\rm e} (r=0)$, upon different central temperature values
(downwards: 4, 6, 8 -- the thickest line -- and 10 keV) is shown. 
All the profiles were calculated at $t_{\rm m} =$ 3 Gyrs, assuming $\beta
= 2/3$ (solid lines) and $\beta = 1$ (dashed lines). 
} \end{figure}

The accretion of infalling material on the cluster potential 
shock heats the protons to an initial temperature $T_{\rm p}^i$, which is
representative of their isotropic Maxwellian velocity distribution. 
The protons equilibrate between themselves in a time shorter than $t_{\rm
eq}$ by a factor $(m_{\rm e}/m_{\rm p})^{1/2} \sim 0.02$, so that they can
be considered to be in equilibrium when the heating event
occurred a time $t_{\rm m} \sim t_{\rm eq}$ in the past. 
The electrons, which are not strongly involved in shock events due to
their negligible mass, achieve a low temperature $T_{\rm e}^i \ll T_{\rm
p}^i$.

Since the total kinetic energy density $U=\frac{3}{2}k(n_{\rm p}T_{\rm p}
+n_{\rm e}T_{\rm e})$ has to be conserved, we have the following
implications:
(i) the local mean gas temperature, $T_{\rm gas}$,
\begin{equation}
T_{\rm gas} = \frac{n_{\rm p} T_{\rm p} + n_{\rm e} T_{\rm e}}{n_{\rm gas}} = 
\frac{T_{\rm p} + 1.21 T_{\rm e}}{2.21}
\end{equation}
is constant with time; (ii) the initial $T_{\rm p}^i$ is about
2.2 times the balanced temperature value, $T_{\rm gas}^i$; 
(iii) $T_{\rm e}$ increases (and $T_{\rm p}$ decreases) with time, with
the energy exchange between protons and electrons driven by the relation
\begin{equation}
\frac{dT_{\rm e}}{dt}=-(n_{\rm p}/n_{\rm e}) \frac{dT_{\rm p}}{dt}.
\end{equation}

Rearranging eqn.(2) in the form:
\begin{equation}
\int_{0}^{T_{\rm e}} \frac{t_{\rm eq}}{T_{\rm p} - {\rm T}_{\rm e}} 
\ d{\rm T}_{\rm e} 
= \int_{0}^{t_{\rm m}} dt,
\end{equation}

and using relation (4), we can solve analytically eqn.(6), obtaining
\begin{eqnarray}
\lefteqn{
\ln \left(\frac{ \sqrt{T_{\rm gas}} + \sqrt{T_{\rm e}} }
{\sqrt{T_{\rm gas}} - \sqrt{T_{\rm e}} } 
\right) -\frac{2}{3} \left(\frac{T_{\rm e}}{T_{\rm gas}} \right)^{3/2} - 2
\left(\frac{T_{\rm e}}{T_{\rm gas}} \right)^{1/2}  = } \nonumber \\ 
 & & \sqrt{\frac{2}{\pi}} \frac{m_{\rm e}}{m_{\rm p}} 
\left(\frac{m_{\rm e} c^2}{k_{\rm B}T_{\rm gas}}
\right)^{3/2} c \sigma_{\rm T} \ln \! \Lambda \ t_{\rm m} n_{\rm gas}.
\end{eqnarray}

Where $T_{\rm e}/T_{\rm gas} < 0.5$, such as in the outer part of 
the temperature profile given a high central temperature and steep density
profile,
we can further expand the logarithmic term in eqn.(6). Then, for
$T_{\rm gas}=$constant, we obtain that $T_{\rm e} \propto n_{\rm
gas}^{2/5}$.

We now parametrise the proton density by a $\beta-$model (Cavaliere \&
Fusco-Femiano 1978), as generally adopted for the description of the ICM
with central density $n_0$ and core radius $r_{\rm c}$,
\begin{equation}
n_{\rm p}(r) = n_0 \left[ 1 + (r/r_{\rm c})^2 \right]^{-3\beta/2},
\end{equation}
and fix the electron density $n_{\rm e} = 1.21 \times n_{\rm p}$.

Results for reference values of $t_{\rm m} = 3$ Gyrs, 
$T_{\rm e} (r=0)$ = 8 keV,
$n_0 = 2.5 \times 10^{-3}$ cm$^{-3}$, $r_{\rm c} = 0.3$ Mpc and $\beta=2/3$,
are shown in Fig. 1. A significant small temperature gradient can remain
at this reference time. In Fig.~(1), we also show
the projected temperature, $<T>$, which is the emissivity-weighted
$T_{\rm e}$: 
\begin{equation} 
<T> = \int_{b^2}^\infty
\frac{\epsilon(r) \ T_{\rm e}(r) \ dr^2}{\sqrt{r^2-b^2}}/
\int_{b^2}^\infty \frac{\epsilon(r)\ dr^2}{\sqrt{r^2-b^2}},
\end{equation}
where the volume emissivity $\epsilon$ is integrated along the line
of sight at projected radius $b$.
It is more appropriate to compare with observed values and
is shown to decrease more significantly than $T_{\rm e}(r)$.

\begin{figure}
\psfig{figure=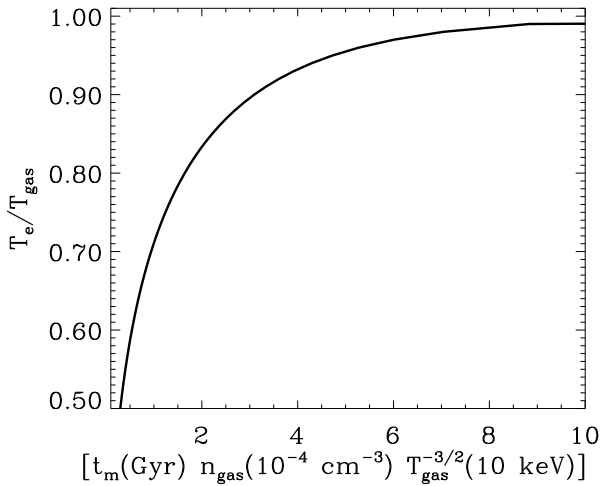,width=0.5\textwidth,angle=0}
\caption{The dependent variable $T_{\rm e}/T_{\rm gas}$ at first member in
eqn.(7) is plotted versus the independent variable present in the second
member. 
} \end{figure}

The dependence of the temperature gradient $T_{\rm e}(r)$ on $t_{\rm m}$,
$\beta$ and the central value of $T_{\rm e}$, $T_{\rm e} (r=0)$, is shown
in Fig.~2 and 3.
Note that the temperature gradient steepen when there is (i) a higher
energy per particle (i.e. when a larger central $T_{\rm e}$ is
considered); (ii) a greater value for $\beta$, so that the gas density
becomes steeper; or, of course, (iii) the timescale is shorter.

In Fig.~4, we show how $T_{\rm e}/T_{\rm gas}$ varies with $t_{\rm m}
n_{\rm gas} T_{\rm gas}^{-3/2}$. A 10 per cent difference in $T_{\rm e}$,
with respect to $T_{\rm gas}$, requires that $n_{\rm gas} < 3.09 \times
10^{-4} t_{\rm m}^{-1} T_{\rm gas}^{3/2} {\rm cm}^{-3}$, where $t_{\rm m}$
is in Gyrs and $T_{\rm gas}$ in units of 10 keV. We note that the outer
densities detectable by the ROSAT {\it PSPC} in deep cluster images are of
order of about $10^{-4} {\rm cm}^{-3}$.

\subsection{The presence of a gradient in the gas temperature}

\begin{figure}
\psfig{figure=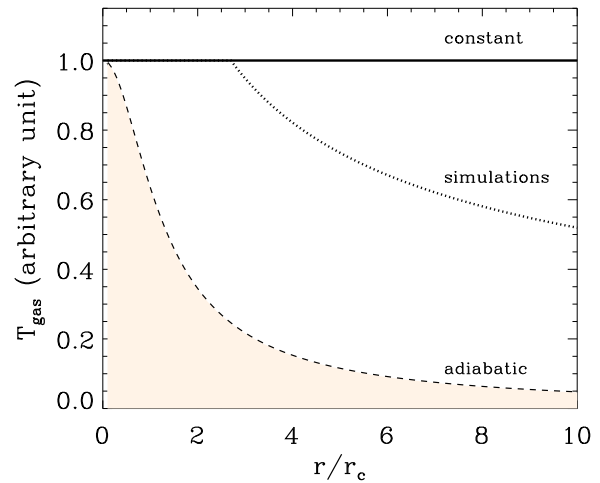,width=0.5\textwidth,angle=0}
\caption{The temperatures profiles discussed in Sect.~2.1 are 
plotted here, assuming (i) a ratio $r_{\rm
c}/r_{200}$ of 0.15 (cf. table~1 in Navarro et al. 1996) for the dotted
profile from the N-body simulations, and (ii) a $\beta$-model with
$\beta=2/3$ for the adiabatic profile (dashed line). The shadowy region
indicates where the gradients cause the ICM to be convectively
unstable. } \end{figure}

Above, we have considered the case of $T_{\rm gas}(r) =$ constant.

However, recent N-body simulations of clusters of galaxies (Navarro
et al. 1996, Evrard et al. 1996) have shown that $T_{\rm gas}$ is almost
constant in the range 0.1--0.4 $r/r_{200}$ (where $r_{200}$ indicates the
radius at which the cluster mean density is 200 times the critical
density value) and slightly decreases as $r^{-1/2}$ towards the outer
virialized part. Such a gradient is not unstable to convective mixing,
from the adiabatic condition [cf. eqn.(1) and Fig.~5]:
\begin{equation}
\frac{dT_{\rm gas}}{dr} > \frac{d \left(n_{\rm gas}^{2/3} \right)}{dr} \ .
\end{equation}

We can use eqn.(7) (obtained under the condition of $T_{\rm gas}$
constant with time) to investigate the more extreme electron temperature
gradient when the gas is adiabatic. We obtain a behavior of $T_{\rm e}$
almost coincident with the $T_{\rm gas}$ profile shown in Fig.~5, for
every $t_{\rm m}$ considered.

\section{The case of A2163} 

A long Coulomb equilibration timescale between electrons and protons has
been suggested by Markevitch et al. as one explanation for the steep
negative temperature profile observed in {\it ASCA} observations of A2163.
We now try to test this possibility by comparing our
calculated profile with the deprojected temperature in Abell 2163
(Markevitch et al. 1996; Elbaz et al. 1995; Fig.~6). In particular, we
find that it is impossible to reconcile the theoretical temperature
gradients with the observed one in the outer parts of the cluster, for
any reasonable $t_{\rm m}$ (even for
1 Gyr the disagreement is about of 5.4 $\sigma$), if the final state for the
ICM is isothermal. Only if we make the {\it ad hoc} assumption that the ICM
is adiabatic beyond about one Mpc can we obtain some agreement (cf.
dashed lines in Fig.~6).

\begin{figure}
\psfig{figure=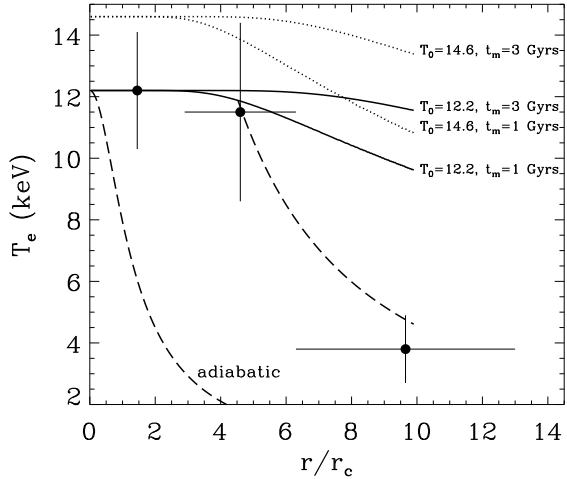,width=0.5\textwidth,angle=0}
\caption{The temperature profile of A2163 from the analysis by Markevitch
et al.
(1996).  The model profiles are calculated using as input the values in
Elbaz et al. (1995): $n_0=6.65 \times 10^{-3}$ cm$^{-3}$, $\beta=0.62$.
$T_{\rm e}(r=0)$ is considered both equal to 12.2 (solid lines; Markevitch
et al.) and 14.6 (dotted lines; Elbaz et al.) keV. The results from an
assumed isothermal profile clearly fail at the outer radius. 
The two adiabatic profiles (dashed lines) are calculated at $r/r_{\rm
c}$ equal to 0 and 4.5 via eqn.(1) and with $\gamma=5/3$. 
From the discussion in
Sect.~2.1, they do not depend significantly on $t_{\rm m}$.} 
\end{figure}

We now consider whether the gas can be considered to be in hydrostatic
equilibrium since the last plausible merger (a few billion years). 
The time
required for a sound wave to cross a cluster is obtained by the relation
$v_{\rm S}^2= dP_{\rm gas}/d\rho_{\rm gas}$, where $P_{\rm gas}$ and
$\rho_{\rm gas}$ are the ICM pressure and density, respectively. Solving
this relation with respect to time, we get $t_{\rm S} \propto r
\ T_{\rm gas}^{-0.5} $, that does not depend significantly on any
temperature (density) gradient. Using the values in Fig.~6 for A2163, we
calculate that $t_{\rm S}$ is $0.6\pm0.6$, $2.1\pm0.8$ and $7.7\pm2.9$
Gyrs at 0.45, 1.43 and 2.99 Mpc, respectively (the errors are $1\sigma$
deviation obtained by propagation on $r$ and $T_{\rm gas}$
uncertainties). A critical condition for the hydrostatic assumption
appears above 1.4 Mpc ($\sim 5 r_{\rm c}$), where the steep temperature
profile requires that a merger occurred less than about 3 Gyrs ago, 
if the steepness is due to the Coulomb process. Then, 
the cluster appears unrelaxed in its outer parts, with the subsequent
presence of gas bulk motions, whose pressure is not accounted for by the
measured ICM temperature in the last spatial bin of Fig.~6 (see also the
discussion in Markevitch et al. 1996). We conclude that in regions of
$(r,t_{\rm m})$ space where the electron temperature is significantly
below the proton temperature then it is likely that hydrostatic
equilibrium has broken down.

Another timescale relevant to the thermalization of the ICM is the
ionization equilibrium time of the abundant elements, $t_{\rm ion}$. The
emission lines of the highly ionized Fe atoms are the strongest features
detected in numerous X-ray spectra of clusters of galaxies. Using the
fitted parameters in Shull \& Van Steenberg (1982) for the collisional
ionization rate coefficients $C_{\rm Fe}$ of the hydrogenic FeXXVI ions,
we get $t_{\rm ion} = (C_{\rm Fe} \ n_{\rm e})^{-1}
\sim 0.1-10$ Gyrs in the range $0 < r/r_{\rm c} < 10$ for the
density profiles considered above (i.e. $\beta = 2/3$ and 1, 
$T_{\rm e} = 8$ keV).
In the case of A2163, $t_{\rm ion}$ is of the order of a few tenth of Gyr,
and less than 0.4 Gyr, which is shorter than any other timescale discussed
here.

Finally, we note that the thermal conduction timescale on which
temperature gradients are erased are much longer than the other
timescales considered here by more than one order of magnitude.

\section{CONCLUSIONS}

In this paper, we have shown that electron--proton Coulomb interactions in
ICM become inefficient in reaching the equipartition, steepen the
electron temperature gradient significantly, only if (i) the
energy per particle is high, (ii) the gas density profile is steep, and
(iii) the time elapsed since the last merger, or proton heating event, is
very short, i.e. $n_{\rm gas} < 3.09 \times 10^{-4} (t_{\rm m}/ 1
{\rm Gyr})^{-1} (T_{\rm gas}/ 10 {\rm keV})^{3/2} {\rm cm}^{-3}$.
Local conservation of energy means that the proton
temperature drops as the electron temperature rises. 
Regions of clusters where a large
disequilibrium occurs are likely to both be out of hydrostatic 
equilibrium and to have low X-ray emission.

Similar conclusions are also presented by Fox \& Loeb (1997), who
consider Coulomb interactions in a plasma accreting on a cluster in the
framework of the spherical self-similar model.
  
When applied to A2163, for which Markevitch et al. (1996) report a steep
drop in electron temperature in the outer parts of the cluster, we find
that we cannot reproduce the profile from gas in hydrostatic equilibrium,
without requiring the mean gas temperature to drop sharply beyond about
one Mpc.

\section*{ACKNOWLEDGEMENTS} We thank Paul Nulsen for discussion, 
and the anonymous referee for suggestions on improving this paper. 
SE acknowledges support from a PPARC 
studentship and a Cambridge European Trust Grant and ACF the support 
of the Royal Society.

\end{document}